\documentclass[12pt]{iopart}

\usepackage{graphicx}
\usepackage{dcolumn}
\usepackage{bm}
\usepackage{citesort}

\begin{document}

\title{Thermalization of a strongly interacting 1D Rydberg lattice gas}

\author{B. Olmos$^1$, M. M\"uller$^2$ and I. Lesanovsky$^3$}
\address{$^1$Instituto 'Carlos I' de F\'{\i}sica Te\'orica y Computacional
and Departamento de F\'{\i}sica At\'omica, Molecular y Nuclear,
Universidad de Granada, E-18071 Granada, Spain}
\address{$^2$Institute for
Theoretical Physics, University of Innsbruck, and Institute for
Quantum Optics and Quantum Information of the Austrian Academy of
Sciences, Innsbruck, Austria}
\address{$^3$Midlands Ultra Cold Atom Research Centre - MUARC, The University of Nottingham, School of Physics and Astronomy, University Park, Nottingham NG7 2RD, United Kingdom}
\ead{igor.lesanovsky@nottingham.ac.uk}
\date{\today}

\begin{abstract}\label{txt:abstract}
When Rydberg states are excited in a dense atomic gas the mean number of excited atoms reaches a stationary value after an initial transient period. We shed light on the origin of this steady state that emerges from a purely coherent evolution of a closed system. To this end we consider a one-dimensional ring lattice, and employ the perfect blockade model, i.e. the simultaneous excitation of Rydberg atoms occupying neighboring sites is forbidden. We derive an equation of motion which governs the system's evolution in excitation number space. This equation possesses a steady state which is strongly localized. Our findings show that this state is to a good accuracy given by the density matrix of the microcanonical ensemble where the corresponding microstates are the zero energy eigenstates of the interaction Hamiltonian. We analyze the statistics of the Rydberg atom number count providing expressions for the number of excited Rydberg atoms and the Mandel Q-parameter in equilibrium.
\end{abstract}

\pacs{32.80.Ee,67.85.-d, 32.80.Rm, 42.50.Nn}
\maketitle
\section{Introduction}
In Rydberg states of alkali atoms the valence electron occupies a high lying orbit and is only loosely bound to the ionic core \cite{Gallagher84}. This large displacement between the charges ($\sim 100\, \mathrm{nm}$) gives rise to a large atomic polarizability \cite{Mohapatra08} and strong interatomic interactions. This has been impressively demonstrated in an ultracold gas where atoms have been laser excited to Rydberg states from their electronic ground state \cite{Heidemann07}. The strong state-dependent interaction forbids the simultaneous excitation of two atoms when they are too close. This effect - the Rydberg blockade - has potential application in quantum information processing \cite{Jaksch00,Urban08,Gaetan08,Mueller09,Wilk09,Isenhower09}, the preparation of many-particle states \cite{Schachenmayer09} in optical lattices and the creation of non-classical light sources \cite{Lukin01,Pohl09}.

A coherently laser-driven gas of Rydberg atoms constitutes an ideal specimen to study the complex many-body dynamics of a strongly interacting quantum system. The underlying Hamiltonian can be formulated as a spin model \cite{Sun08,Raitzsch08,Weimer08,Olmos09,Olmos09-3,Schachenmayer09} whose properties are tunable through the laser parameters and the choice of particular Rydberg states. The experimental realization of such a system has been achieved in several labs and the detection of Rydberg atoms can be carried out with high efficiency. Hence, observables such as the number of Rydberg atoms can be recorded with high accuracy. A typical experiment starts with all atoms being initialized in the ground state. The laser which couples the ground to the Rydberg state is turned on for a certain time $t$ after which the number of Rydberg atoms is detected \cite{Singer04,Tong04,Heidemann07}.
\begin{figure}\center
\includegraphics[width=.7\columnwidth]{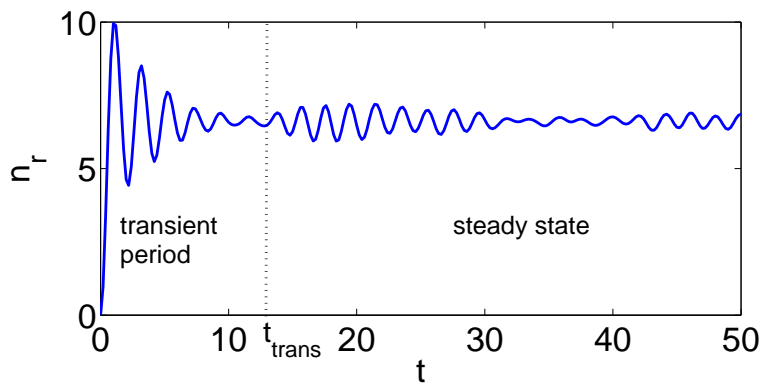}\caption{Number of Rydberg atoms as a function of time when initially no Rydberg atom is excited. One observes an initial (transient) phase with large contrast oscillations followed by a steady state. These features do not depend on the actual details of the system, i.e. the exact atomic interaction, the atomic arrangement and the boundary conditions. The data here is shown for a ring lattice with $25$ sites. Each site is occupied by the same number of atoms $N_0$ and the atoms are resonantly coupled to a Rydberg state by a laser with Rabi frequency $\Omega_0$. The interaction strength between neighboring Rydberg atoms is taken to be infinite. The next-nearest neighbor interaction is zero. The time is given in units of the inverse collective Rabi frequency $\Omega^{-1}$.}
\label{fig:generic_rydberg_number}
\end{figure}
The resulting signal is determined by the interplay between the laser driving and the strong state-dependent interaction \cite{Heidemann07}.
The typical appearance of the temporal evolution of the Rydberg number $n_\mathrm{r}$ obtained from theoretical studies \cite{Sun08,Olmos09,Weimer08,Schachenmayer09} is depicted in fig. \ref{fig:generic_rydberg_number}.
At small times ($t\leq t_\mathrm{trans}$) $n_\mathrm{r}$ rises first quadratically in $t$ showing subsequently a number of oscillations with high contrast. After this transient period the oscillations diminish and $n_\mathrm{r}$ approaches a constant value around which it performs small amplitude fluctuations which decrease with increasing system size. This general appearance is independent of the exact details of the system, e.g. the actual arrangement of the atoms and the boundary conditions. In experiments which are carried out in disordered gases the individual oscillations might not be observable since averaging over many experimental realizations also means averaging over many different spatial distributions of the atoms \cite{Raitzsch08}. Each of these distributions gives rise to a different shape of the oscillating features in fig. \ref{fig:generic_rydberg_number} and eventually the oscillations are washed out.

The steady state shown in fig. \ref{fig:generic_rydberg_number} is the result of a purely \emph{coherent dynamics of a closed system} and does not come about due to dissipation stemming from the coupling to an external bath. In experiments a fundamental cause of dissipation would be radiative decay of atoms which typically occurs on a timescale $\tau_\mathrm{rad}\sim 10-100\, \mu\mathrm{s}$. This is long compared to the time it takes to carry out the experiment $\tau_\mathrm{exp}\sim\, 1 \mu\mathrm{s}$. Consequently, it is well-justified to neglect radiative decay in theoretical models of gases of Rydberg-excited atoms when the time-evolution is restricted to time intervals smaller than $\tau_\mathrm{rad}$. Since any numerical treatment can only encompass a finite number of degrees of freedom, there is always a revival time $t_\mathrm{rev}$ after which the system returns to its initial state. The steady state which is observed in fig. \ref{fig:generic_rydberg_number} can thus be itself only a transient effect. However, in practice we have $\tau_\mathrm{rad}\ll t_\mathrm{rev}$ and thus the common theoretical models which consider only coherent dynamics are invalid for such long times. Thus, the steady state we are referring to in this work exists in the time interval $t_\mathrm{trans}<t<\tau_\mathrm{rad}$.

The purpose of this work is to study the origin of this steady state which occurs as result of a purely coherent dynamics of a closed system. This subject is closely related to the discussion about how and why a closed system which is prepared in a pure state actually thermalizes. I.e. why and how does such system assume a state in which the mean values of macroscopic observables are stationary and why can these mean values be calculated from the microcanonical ensemble? These questions are of fundamental interest \cite{Deutsch91,Srednicki94} and have been investigated in a number of systems \cite{Kinoshita06,Rigol07,Kollath07,Hofferberth07,Rigol08}.

In the present work it is particularly insightful to study the evolution of the atomic gas in excitation number space, i.e. between subspaces of the system's Hilbert space which contain the same number of Rydberg atoms. Throughout, we employ the \emph{perfect blockade} model of a Rydberg gas which has been extensively discussed in Refs. \cite{Sun08,Olmos09}: Two-level atoms (ground state $\left|g\right>$, Rydberg state $\left|r\right>$) are confined to a ring lattice with $N$ sites. Each site contains the same number ($N_0$) of atoms, which occupy a single spatial mode and no tunneling between adjacent sites occurs. I.e. the external degrees of freedom are frozen out. A laser with Rabi frequency $\Omega$ couples $\left|r\right>$ and $\left|g\right>$, resonantly. The interaction between Rydberg atoms is so strong that the interaction energy of two atoms occupying the same site or adjacent sites is infinite \footnote{In practice this means that the interaction energy has to be much larger than $\Omega$.}. The interaction between Rydberg atoms being separated by two or more sites is zero.

The physical degrees of freedom are the two internal states of the atoms located at individual sites, i.e. their measurement eventually determines $n_\mathrm{r}$. The state-dependent interaction in conjunction with the laser coupling, however, leads to such a strong mixing that perturbation theory in these physical degrees of freedom becomes meaningless. Instead, the eigenstates of the system are collective, delocalized excitations which contain no well-defined number of Rydberg atoms. When studying the evolution in excitation number space under the action of the laser, we find that the strong interaction causes quasi-random couplings between regions of the Hilbert space which contain a different (and well-defined) number of Rydberg atoms. This randomness allows us to derive an effective equation for the time-evolution of the probability of being in a subspace with certain fixed excitation number. The resulting equation possesses a steady state which solely depends on the dimension of the excitation number subspaces. A comparison to the results obtained from the numerically exact propagation of the Schr\"odinger equation shows good agreement.

\section{Hamiltonian}
\subsection{Formulation in the physical degrees of freedom}
We consider atoms trapped in a ring-shaped one-dimensional lattice with $N$ sites. We will see, that despite its simplicity this model offers interesting insights into the dynamics of laser-driven Rydberg gases which are usually carried out in a gas cloud. The Hamiltonian is given by
\begin{eqnarray}
  H=H_0+H_\mathrm{int}=\Omega\sum_{k=1}^N \sigma^{(k)}_x + \frac{\beta}{4}\sum_{k>l}^N \frac{ \left(1+\sigma_z^{(k)}\right)\left(1+\sigma_z^{(l)}\right)}{|k-l|^\gamma}\label{eq:working_hamiltonian}
\end{eqnarray}
where $\sigma_x^{(k)}$ and $\sigma_z^{(k)}$ are the usual Pauli spin matrices acting on the internal state of the atom located at site $k$ and $\sigma_x^{(N+1)}=\sigma_x^{(1)}$. The first part ($H_0$) describes the interaction of the atoms with the laser within the rotating-wave approximation. $H_\mathrm{int}$ accounts for interaction between Rydberg atoms located different sites. The atomic states are identified as $\left|g\right>_k\equiv \left|\downarrow\right>_k$ and $\left|r\right>_k\equiv \left|\uparrow\right>_k$ and $\left(1+\sigma_z^{(k)}\right)/2$ is the projector on the Rydberg state at site $k$. We are interested here in atomic Rydberg states that interact via the van-der-Waals interaction. In this case we have $\gamma=6$ and the parameter $\beta=C_6/a^6$ determines the interaction strength with $C_6$ being the van-der-Waals coefficient \cite{Singer05} and $a$ the lattice spacing. The transition $\left|g\right>_k\rightarrow\left|r\right>_k$ is resonantly (zero detuning) driven with the Rabi frequency $\Omega$. In case of a single atom per site $\Omega$ is equal to the single atom Rabi frequency $\Omega_0$. If each site is occupied by the same number ($N_0\gg 1$) of atoms $\Omega$ corresponds to the collective Rabi frequency $\sqrt{N_0}\Omega_0$. In this case the laser couples the product state $\left|g\right>_k \equiv \left[\left|g\right>_k\right]_1\otimes ... \otimes \left[\left|g\right>_k\right]_{N_0}$ to the superatom state $\left|r\right>_k\equiv \mathcal{S}\left\{\left[\left|r\right>_k\right]_1\otimes \left[\left|g\right>_k\right]_2\otimes ... \otimes \left[\left|g\right>_k\right]_{N_0}\right\}$ ($\mathcal{S}$ is the symmetrization operator). The derivation and the validity of Hamiltonian (\ref{eq:working_hamiltonian}) is thoroughly discussed in Refs. \cite{Olmos09,Schachenmayer09}.
\begin{figure}\center
\includegraphics[width=0.7\columnwidth]{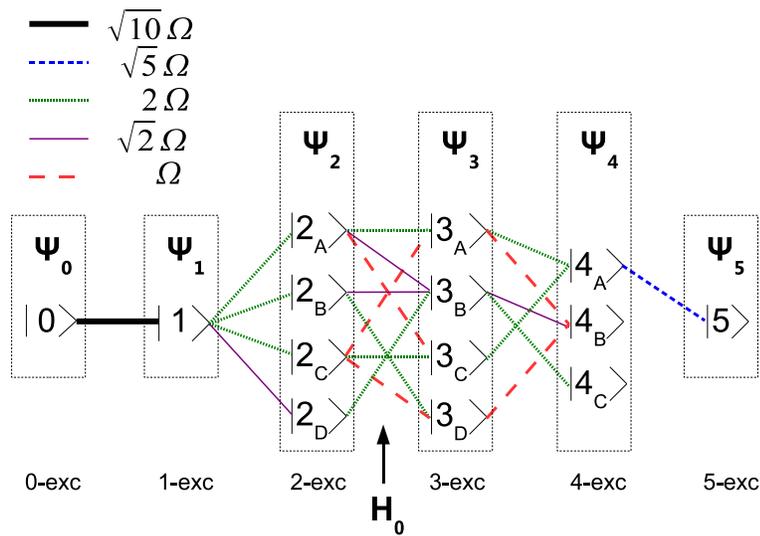}
\caption{Irreducible graph for a lattice with $N=10$ sites in which the time-evolution takes place. Each column contains states with the same number of Rydberg excitations. The laser (Hamiltonian $H_0$) only couples adjacent subspaces. The coupling strength (transition probability) between the individual states is encoded in the colors and line style.}\label{fig:graph}
\end{figure}

We are now interested in the regime of the so-called  \emph{perfect blockade} which was introduced in Refs. \cite{Sun08,Olmos09}. In this limit we replace the van-der-Waals interaction by an interaction potential whose value is $\beta\gg\Omega$ for nearest neighbors and $0$ for atoms which are further apart. This means that we effectively consider the case in which $\gamma$ in eq. (\ref{eq:working_hamiltonian}) is infinite. The perfect blockade model is then valid provided that $1\gg \Omega/\beta = a^6\Omega/C_6 \gg 1/64$, i.e. $\Omega$ has to be much smaller than the nearest neighbor interaction but at the same time much larger than the next nearest neighbor interaction. In practice this means $\beta\simeq 10\Omega$. For such a choice the further results, which will be obtained under the assumption of the perfect blockade, are virtually indistinguishable from those obtained for van-der-Waals interacting atoms. Such good agreement was also reported in Ref. \cite{Sun08} and is a speciality of the one-dimensional nature of our system; for comparison: In two dimensions the perfect blockade model was valid provided that $1\gg \Omega/\beta \gg 1/8$ which is actually difficult to satisfy. The perfect blockade approach has the advantage that certain aspects of the system, e.g. its steady state and derived quantities, can be calculated analytically. Within the \emph{perfect blockade} the only states $\left|\phi\right>$ which participate in the dynamics are those which satisfy
\begin{eqnarray}
  \left(\lim_{\gamma\rightarrow\infty} H_\mathrm{int}\right)\left|\phi\right>=0.\label{eq:non_interacting_condition}
\end{eqnarray}
The restriction to this set of states is the manifestation of the strong interaction among the Rydberg atoms. The states $\left|\phi\right>$ can contain any number of Rydberg excitations between $0$ and $n_\mathrm{max}$, which is the next integer smaller than $N/2$.

The laser Hamiltonian $H_0$ can only couple states whose excitation number differs by one. Moreover, due to the differences in the spatial distribution of the excitations only certain states are connected by the laser. A convenient way to illustrate the coupling is a graph as shown in fig. \ref{fig:graph}. Here vertices in the same column contain the same number of Rydberg atoms and adjacent columns are connected by the laser Hamiltonian $H_0$. A similar approach to the representation of an interacting many particle system was employed by Altshuler \textit{et al.} in Ref. \cite{Altshuler97}. Here a system of interacting fermions was represented as a graph whose vertices are represented by Fock states. Coupling between these states was caused by the interaction. In our case the interaction determines the vertices and the single particle Hamiltonian ($H_0$) determines their coupling. Fig. \ref{fig:graph} shows the fully reduced graph for $N=10$ where only states which are invariant under reversal ($k\rightarrow N-k+1$) and cyclic shifts ($k\rightarrow k+1$) of the sites of the ring lattice are considered (see Ref. \cite{Olmos09}). The state in which all atoms are in the ground state, i.e. $\left|0\right>=\prod_{k=1}^N\left|g\right>_k$, is contained in this fully symmetric set. Solving the Schr\"odinger equation of Hamiltonian (\ref{eq:working_hamiltonian}) with this (vacuum) state as initial state and calculating the mean number of Rydberg atoms
\begin{eqnarray}
  n_\mathrm{r}(t)=\left<\Psi(t)\right|\sum_{k=1}^N \frac{1+\sigma_z^{(k)}}{2}\left|\Psi(t)\right>\label{eq:ryd_num_phys_DOF}
\end{eqnarray}
with $\left|\Psi(t)\right>=\exp(-iHt)\left|0\right>$
eventually gives rise to the data presented in fig. \ref{fig:generic_rydberg_number}.

\subsection{Hamiltonian in excitation number space}
Due to the strong interaction a formulation of the problem in terms of the physical degrees of freedom (localized atoms) seems disadvantageous. Instead, it appears more natural to base the discussion on the graph depicted in fig. \ref{fig:graph}. Initially the system is localized on the leftmost vertex of the graph, i.e. it resides in the vacuum state $\left|0\right>=\prod_{k=1}^N\left|g\right>_k$. Once the laser is turned on, coupling to neighboring vertices is established and the propagation of population through the graph sets in. Eventually, the mean Rydberg number is determined by the probability $\rho_m(t)$ of the system to be in the subspace containing $m$ excitations:
\begin{eqnarray}
  n_\mathrm{r}(t)=\sum_{m=0}^{n_\mathrm{max}} m\, \rho_m(t)\label{eq:ryd_num_graph}.
\end{eqnarray}
$\rho_m(t)$ is hereby the probability density derived from $\left|\Psi(t)\right>$ and integrated over each column of the graph. Note, that expressions (\ref{eq:ryd_num_phys_DOF}) and (\ref{eq:ryd_num_graph}) are equivalent.

In order to make the excitation number spaces explicitly appear in the equations, we use the following matrix representation of the wave function:
\begin{eqnarray}
\mathbf{\Psi}=\left(
                \begin{array}{c}
                  \vdots \\
                  \mathbf{\Psi}_{m-1} \\
                  \mathbf{\Psi}_m \\
                  \mathbf{\Psi}_{m+1} \\
                  \vdots \\
                \end{array}
              \right)=\mathbf{\Psi}_{0}\oplus ... \oplus \mathbf{\Psi}_m \oplus ... \oplus \mathbf{\Psi}_{n_\mathrm{max}}.
\end{eqnarray}
The vectors $\mathrm{\Psi}_m$ are the projections of the wave function onto the space with $m$ excitations and contain $\mathrm{dim}_m$ components which are labeled by the indices $\alpha_{m}=1,...,\mathrm{dim}_m$, i.e. $\mathrm{dim}_m$ is the number of possibilities of placing $m$ Rydberg atoms on the ring that are compatible with the constraint (\ref{eq:non_interacting_condition}). The probabilities $\rho_m$ are defined by
\begin{eqnarray}
  \rho_m=\mathbf{\Psi}^\dagger_m\mathbf{\Psi}_m.
\end{eqnarray}
In this representation the Hamiltonian becomes
\begin{eqnarray}
  H=\left(
      \begin{array}{ccccc}
        0 & \ddots & 0 & 0 & 0 \\
        \ddots & 0 & \mathcal{C}_{n-1,n} & 0 & 0 \\
        0 & \mathcal{C}^\dagger_{n-1,n} & 0 & \mathcal{C}_{n,n+1} & 0 \\
        0 & 0 & \mathcal{C}^\dagger_{n,n+1} &0 & \ddots \\
        0 & 0 & 0 & \ddots & 0 \\
      \end{array}
    \right)\label{eq:hamiltonian}
\end{eqnarray}
where the operators $\mathcal{C}_{n-1,n}$ and $\mathcal{C}^\dagger_{n-1,n}$ connect the subspaces which contain $n-1$ and $n$ Rydberg atoms. Each of the subspaces contains $\mathrm{dim}_n$ states.  Hence, $\mathcal{C}_{n,n+1}$ is a $\mathrm{dim}_{n} \times \mathrm{dim}_{n+1}$-matrix. The block structure of the Hamiltonian (\ref{eq:hamiltonian}) is a direct consequence of the property of the laser Hamiltonian to couple only subspaces whose excitation number differs by one.

The projection operator onto the space containing $n$ Rydberg atoms is given by $\left|n\right>\left<n\right|$, and we can define $\mathrm{\Psi}_n=\left<n\mid \Psi \right>$. This allows us to rewrite Hamiltonian (\ref{eq:hamiltonian}) more compactly as
\begin{eqnarray}
 \fl H=\sum_{n,m=0}^{n_\mathrm{max}} \left|m\right>\left<m\right|H\left|n\right>\left<n\right|=\sum_{n=0}^{n_\mathrm{max}-1}\left[ \mathcal{C}_{n,n+1}\left|n\right>\left<n+1\right|+\mathcal{C}^\dagger_{n,n+1}\left|n+1\right>\left<n\right|\right].
\end{eqnarray}

\section{Time-evolution in the excitation number subspace}
\subsection{Time-evolution of the projection operators}
Our goal is to calculate the time-evolution of the projection operators $\left|m\right>\left<m\right|$. This will eventually enable us to calculate the quantities $\rho_m(t)$. To this end we consider the Heisenberg equation of motion
\begin{eqnarray}
\partial_t \left|m\right>\left<m\right|=i\left[H,\left|m\right>\left<m\right|\right]
\end{eqnarray}
which can be formally integrated to yield
\begin{eqnarray}
  \left|m\right>\left<m\right|_t-\left|m\right>\left<m\right|_0=i\int_0^t d t^\prime \left[H,\left|m\right>\left<m\right|_{t^\prime}\right].
\end{eqnarray}
For small times $\tau$ we obtain up to second order
\begin{eqnarray}
  \left|m\right>\left<m\right|_\tau-\left|m\right>\left<m\right|_0=i\tau\left[H,\left|m\right>\left<m\right|_0\right]
  -\frac{\tau^2}{2}\left[H,\left[H,\left|m\right>\left<m\right|_0\right]\right].\label{eq:projector_small_t}
\end{eqnarray}
With $\left|m\right>\left<m\right|_0\equiv\left|m\right>\left<m\right|$ the first commutator evaluates to
\begin{eqnarray}
  \left[H,\left|m\right>\left<m\right|\right]&=&\mathcal{C}_{m-1,m} \left|m-1\right>\left<m\right|+ \mathcal{C}^\dagger_{m,m+1}\left|m+1\right>\left<m\right|\nonumber\\
  &&-\mathcal{C}_{m,m+1}\left|m\right>\left<m+1\right|-\mathcal{C}^\dagger_{m-1,m}\left|m\right>\left<m-1\right|\\
  &=&\mathcal{C}^\dagger_{m,m-1} \left|m-1\right>\left<m\right| -\mathcal{C}_{m,m-1}\left|m\right>\left<m-1\right|\nonumber\\
  &&+\mathcal{C}^\dagger_{m,m+1}\left|m+1\right>\left<m\right|-\mathcal{C}_{m,m+1}\left|m\right>\left<m+1\right|.
\end{eqnarray}
where we have used $\mathcal{C}_{m-1,m}=\mathcal{C}^\dagger_{m,m-1}$. The second double-commutator becomes
\begin{eqnarray}
\fl\frac{1}{2} \left[H,\left[H,\left|m\right>\left<m\right|\right]\right]&=&\left[\mathcal{C}_{m,m+1}\mathcal{C}^\dagger_{m,m+1}+\mathcal{C}_{m,m-1}\mathcal{C}^\dagger_{m,m-1}\right]\left|m\right>\left<m\right|\\
\fl&&-\mathcal{C}^\dagger_{m,m-1}\mathcal{C}_{m,m-1}\left|m-1\right>\left<m-1\right|-\mathcal{C}^\dagger_{m,m+1}\mathcal{C}_{m,m+1}\left|m+1\right>\left<m+1\right|\nonumber\\
\fl&&-\mathcal{C}^\dagger_{m-1,m}\mathcal{C}_{m,m+1}\left|m-1\right>\left<m+1\right|-\mathcal{C}^\dagger_{m,m+1}\mathcal{C}_{m,m-1}\left|m+1\right>\left<m-1\right|\nonumber\\
\fl&&+\frac{1}{2}\left[\mathcal{C}^\dagger_{m-1,m-2}\mathcal{C}^\dagger_{m,m-1}\left|m-2\right>\left<m\right|+\mathcal{C}_{m,m-1}\mathcal{C}_{m-1,m-2}\left|m\right>\left<m-2\right|\right.\nonumber\\
\fl&&\left.+\mathcal{C}^\dagger_{m+1,m+2}\mathcal{C}^\dagger_{m,m+1}\left|m+2\right>\left<m\right|+\mathcal{C}_{m,m+1}\mathcal{C}_{m+1,m+2}\left|m\right>\left<m+2\right|\right]\nonumber.
\end{eqnarray}

\subsection{Effective equation of motion for $\rho_m$}
Using eq. (\ref{eq:projector_small_t}) and the definition $\rho_m=\mathbf{\Psi}^\dagger_m\mathbf{\Psi}_m=\left<\Psi\mid m\right>\left<m\mid\Psi\right>$ we find
\begin{eqnarray}
\fl  \rho_m(\tau)-\rho_m(0)=i\tau\left<\Psi\right|\left[H,\left|m\right>\left<m\right|\right]\left|\Psi\right>
  -\frac{\tau^2}{2}\left<\Psi\right|\left[H,\left[H,\left|m\right>\left<m\right|\right]\right]\left|\Psi\right>.\label{eq:equation_for_rho}
\end{eqnarray}
The first commutator contains terms of the form
\begin{eqnarray}
\fl\left<\Psi\mid m-1\right> \mathcal{C}^\dagger_{m,m-1} \left<m\mid \Psi\right>&=&\Psi^\dagger_{m-1} \mathcal{C}^\dagger_{m,m-1} \Psi_{m}\nonumber\\
\fl&=&\sum_{\alpha_{m-1}=1}^{\mathrm{dim}_{m-1}}\sum_{\beta_m=1}^{\mathrm{dim}_{m}} \left[\Psi^\dagger_{m-1}\right]_{\alpha_{m-1}} \left[\mathcal{C}^\dagger_{m,m-1}\right]_{\alpha_{m-1},\beta_m} \left[\Psi_{m}\right]_{\beta_m}\nonumber\\
\fl  &=&\sum_{\alpha_{m-1} \beta_m} \left[\Psi^\dagger_{m-1}\right]_{\alpha_{m-1}} \left[\mathcal{C}_{m,m-1}\right]_{\beta_m,\alpha_{m-1}} \left[\Psi_{m}\right]_{\beta_m}. \label{eq:neglected_term_1}
\end{eqnarray}
Here we have exploited that $\mathcal{C}^\dagger_{m-1,m}$ is a real matrix.

The interaction between the Rydberg atoms manifests itself in the structure of the matrices $\mathcal{C}^\dagger_{m-1,m}$ which were initially constructed in a product basis in which the single atoms constitute the fundamental degrees of freedom. The strong interaction, however, favors collective excitations which are complex superpositions of the single atom excitations. It is thus reasonable to assume that the single atom degrees of freedom are so strongly mixed that the entries of $\mathcal{C}^\dagger_{m-1,m}$ can be regarded as uncorrelated. In this case the expression
\begin{eqnarray}
\Sigma=\sum_{\alpha_{m-1} \beta_m} \left[\Psi^\dagger_{m-1}\right]_{\alpha_{m-1}} \left[\mathcal{C}_{m,m-1}\right]_{\beta_m,\alpha_{m-1}} \left[\Psi_{m}\right]_{\beta_m}
\end{eqnarray}
just becomes a sum of random complex numbers. Its magnitude, i.e. $|\Sigma|$ can be estimated as follows: The components of the wave vector can be approximated by $\left[\Psi_{m}\right]_{\beta_m}\approx (\mathrm{dim})^{-1/2}e^{i\phi_{\beta_m}}$, with $\mathrm{dim}=\sum_{n=0}^{n_\mathrm{max}}\mathrm{dim}_n$ and $\phi_{\beta_m}$ being some phase. With this we can estimate
\begin{eqnarray}
  |\Sigma|\sim \frac{c_{m,m-1}}{\mathrm{dim}}\left|\sum_{\alpha_{m-1} \beta_m}e^{i(\phi_{\beta_m}-\phi_{\alpha_{m-1}})}\right|\approx \frac{c_{m,m-1}}{\mathrm{dim}}\sqrt{\mathrm{dim}_m\,\mathrm{dim}_{m-1}},
\end{eqnarray}
where we have assumed that the complex numbers $e^{i(\phi_{\beta_m}-\phi_{\alpha_{m-1}})}$ are randomly (uniformly) distributed. This allows us to employ the relation $|\sum_{k=1}^N e^{i\alpha_k}|\approx \sqrt{N}$ since in case of randomly distributed $\alpha_k$ we are just dealing with a random walk in two dimensions. The constant $c_{m,m-1}$ relates to the mean value of the entries of the matrix $\mathcal{C}_{m,m-1}$.

The same line of argument holds true for terms stemming from the double commutator which are of the form
\begin{eqnarray}
  \left<\Psi\mid m\right>\mathcal{C}_{m,m+1}\mathcal{C}_{m+1,m+2}\left<m+2\mid\Psi\right>. \label{eq:neglected_term_2}
\end{eqnarray}
Since the entries of the matrices $\mathcal{C}_{m,m+1}$ and $\mathcal{C}_{m+1,m+2}$ are not correlated, their product is again a matrix with randomly distributed elements. The magnitude of these terms can be estimated by employing again the picture of the random walk in two dimensions. The modulus of (\ref{eq:neglected_term_2}) then approximately evaluates to $c_{m,m+1,m+2}\sqrt{\mathrm{dim}_m\,\mathrm{dim}_{m+2}}/\mathrm{dim}$, where $c_{n,m+1,m+2}$ is a constant related to the mean value of the entries of ${\cal C}_{m,m+1}{\cal C}_{m+1,m+2}$.

Qualitatively different, however, are the terms of the form
\begin{eqnarray}
\fl  \left<\Psi\mid m\right>\mathcal{C}_{m,m+1}\mathcal{C}^\dagger_{m,m+1}\left<m\mid\Psi\right>=\sum_{\alpha_{m} \beta_m} \left[\Psi^\dagger_{m}\right]_{\alpha_{m}} \left[\Psi_{m}\right]_{\beta_m} \sum_{\gamma_{m+1}} \left[\mathcal{C}_{m,m+1}\right]_{\alpha_m,\gamma_{m+1}} \left[\mathcal{C}_{m,m+1}\right]_{\beta_m,\gamma_{m+1}},\nonumber
\end{eqnarray}
where the matrix $\mathcal{C}_{m,m+1}$ appears twice. Since the matrix elements of $\mathcal{C}_{m,m+1}$ are uncorrelated only the diagonal elements of the matrix product yield on average a non-zero value, hence
\begin{eqnarray}
  \sum_{\gamma_{m+1}} \left[\mathcal{C}_{m,m+1}\right]_{\alpha_m,\gamma_{m+1}} \left[\mathcal{C}_{m,m+1}\right]_{\beta_m,\gamma_{m+1}}\approx \left[\kappa_{m,m+1}\right]_{\alpha_m}\delta_{\alpha_m,\beta_m}.
\end{eqnarray}
Moreover, since the results cannot depend on the choice of the basis functions spanning a given $m$-excitation subspace, we can say that $\left[\kappa_{m,m+1}\right]_{\alpha_m}=\kappa_{m,m+1}$ and hence
\begin{eqnarray}
\fl  \left<\Psi\mid m\right>\mathcal{C}_{m,m+1}\mathcal{C}^\dagger_{m,m+1}\left<m\mid\Psi\right>&\approx& \kappa_{m,m+1} \sum_{\alpha_{m}} \left[\Psi^\dagger_{m}\right]_{\alpha_{m}} \left[\Psi_{m}\right]_{\alpha_m}\nonumber\\
&=&\kappa_{m,m+1} \mathbf{\Psi}^\dagger_m \mathbf{\Psi}_m=\kappa_{m,m+1} \rho_m(0).\label{eq:dominant_term}
\end{eqnarray}
Estimating $\kappa_{m,m+1}\propto\mathrm{dim}_{m+1}$ the modulus of these (diagonal) terms is proportional to $\mathrm{dim}_{m}\mathrm{dim}_{m+1}/\mathrm{dim}$.

We now return to eq. (\ref{eq:equation_for_rho}). We neglect all terms of the form (\ref{eq:neglected_term_1}) and (\ref{eq:neglected_term_2}) keep only the dominant ones, i.e. those which contain products of the form $\mathcal{C}_{m,m+1}\mathcal{C}^\dagger_{m,m+1}$. By this we obtain
\begin{eqnarray}
  \rho_m(\tau)-\rho_m(0)&\approx& -\tau^2\left[\kappa_{m,m+1}+\kappa_{m,m-1}\right]\rho_m(0)\nonumber\\
&&+\tau^2\left[\kappa_{m-1,m}\rho_{m-1}(0)+\kappa_{m+1,m}\rho_{m+1}(0)\right].
\end{eqnarray}
Here we have used
\begin{eqnarray}
  \sum_{\alpha_{m}} \left[\mathcal{C}_{m,m+1}\right]_{\alpha_m,\beta_{m+1}} \left[\mathcal{C}_{m,m+1}\right]_{\alpha_m,\gamma_{m+1}}\approx \kappa_{m+1,m}\delta_{\beta_{m+1},\gamma_{m+1}},
\end{eqnarray}
from which follows that
\begin{eqnarray}
  \frac{\kappa_{m+1,m}}{\kappa_{m,m+1}}=\frac{\mathrm{dim}_m}{\mathrm{dim}_{m+1}}.
\end{eqnarray}
We can thus make the ansatz $\kappa_{m,m+1}=u_m \mathrm{dim}_{m+1}$ and $\kappa_{m+1,m}=u_m \mathrm{dim}_{m}$ and find, omitting the $t=0$ argument of $\rho_m(0)$,
\begin{eqnarray}
  \rho_m(\tau)-\rho_m&\approx& -\tau^2\left[u_m \mathrm{dim}_{m+1} +u_{m-1} \mathrm{dim}_{m-1}\right]\rho_m\nonumber\\
  &&+\tau^2\left[u_{m-1}\mathrm{dim}_{m}\rho_{m-1}+u_m \mathrm{dim}_m\rho_{m+1}\right].\label{eq:final_equation_for_rho}
\end{eqnarray}
This equation neglects coherent processes which have effectively been eliminated by the neglect of terms of the form (\ref{eq:neglected_term_1}) and (\ref{eq:neglected_term_2}). Thus eq. (\ref{eq:final_equation_for_rho}) cannot be valid for arbitrary small values of $\tau$ as here certainly coherent effects dominate the evolution. Instead, the interval $\tau$ has to be chosen sufficiently large such that terms of the form (\ref{eq:dominant_term}) dominate all other contributions whose importance diminishes due to the summation of complex numbers with random phases. Eq. (\ref{eq:final_equation_for_rho}) is thus a map that propagates the vector  $\mathbf{\rho}(t_0)$ by a 'coarse-grained' timestep $\tau$, i.e. $\mathbf{\rho}(t_0)\rightarrow \mathbf{\rho}(t_0+\tau)$. $\tau$ is thereby chosen much smaller than the typical timescale which governs the evolution of $\rho_m(t)$.

\section{The steady state}
The steady state is defined through
\begin{eqnarray}
  \rho_n^\mathrm{steady}(\tau)-\rho_n^\mathrm{steady}=0,\label{eq:steady_state_condition}
\end{eqnarray}
i.e. it is a fix point of the mapping (\ref{eq:final_equation_for_rho}). The mapping contains the unknown coefficients $u_{m}$ which contain information about  how adjacent excitation subspaces are connected. Fortunately, in order to determine the steady state their knowledge is not necessary. It is only required that $u_{m}\neq 0$, which is always the case. The solution of eq. (\ref{eq:steady_state_condition}) is given by
\begin{eqnarray}
  \rho_n^\mathrm{steady}=\frac{\mathrm{dim}_n}{\mathrm{dim}}.\label{eq:equilibrium_state}
\end{eqnarray}
Thus it is only the number of states contained in a given excitation number subspace that determines the steady state. It is actually possible to calculate the dimension of the subspaces with fixed number of excitations, analytically. This is done by counting all states of the single atom product basis that contain $m$ excited atoms and obey condition (\ref{eq:non_interacting_condition}). The result is
\begin{eqnarray}
 \fl \rho_m^\mathrm{steady}=\frac{1}{\mathrm{dim}}\,\frac{N}{N-m}\left(
 \begin{array}{c}
 N-m \\ m \\
 \end{array}
 \right)
   \qquad&\mathrm{with} &\qquad \mathrm{dim}=\sum_{m=0}^{n_\mathrm{max}}\frac{N}{N-m}\left(
   \begin{array}{c}
   N-m \\ m \\
   \end{array}
   \right).\label{eq:rho_analytical}
\end{eqnarray}
Note, that this result encompasses all possible basis states and not only the fully symmetric ones which constitute the graph (fig. (\ref{fig:graph})). It is interesting to see how $\rho_m^\mathrm{steady}$ behaves in the limit of large $N$, where we have $n_\mathrm{max}=N/2$. It is convenient to introduce the variable $\alpha=m/N$ which is the number of Rydberg atoms divided by the number of sites. Using Stirling's formula we can approximate eq. (\ref{eq:rho_analytical}) by
\begin{eqnarray}\label{eq:rho_stirling}
\rho_m^\mathrm{steady}\propto \frac{1}{\sqrt{2\pi N}}\sqrt{\frac{1-\alpha}{\alpha(1-2\alpha)}}\left(\frac{(1-\alpha)^{1-\alpha}}{\alpha^\alpha(1-2\alpha)^{1-2\alpha}}\right)^N.
\end{eqnarray}
This function has a very pronounced peak and, for large $N$, we can approximate $\rho_m^\mathrm{steady}(\alpha)$ by a Gaussian. The position of the maximum of the function is the solution of the equation
\begin{eqnarray}
\ln{\alpha}+\ln{(1-\alpha)}-2 \ln{(1-2\alpha)}-\frac{1}{N(1-2\alpha)}+\frac{1}{2N\alpha(1-\alpha)}=0,
\end{eqnarray}
where $\ln(x)$ is the natural logarithm. Since the number of sites $N$ is taken to be very large, we can neglect the last two terms, and then obtain that the function (\ref{eq:rho_stirling}) assumes its maximum at
\begin{eqnarray}
  \alpha_\mathrm{max}=\frac{1}{2}\left[1-\frac{1}{\sqrt{5}}\right]\approx 0.276.
\end{eqnarray}
Around this peak value the squared width of $\rho_m^\mathrm{steady}(\alpha)$ is given by
\begin{eqnarray}
  \sigma^2_\alpha=\frac{1}{5\,\sqrt{5} N}
\end{eqnarray}
Hence, for large $N$ the probability distribution in particle number space is strongly peaked with an overwhelming weight on $\alpha_\mathrm{max}$. The mean number of Rydberg atoms in the steady state is thus expected to be $\bar{n_\mathrm{r}}=\alpha_\mathrm{max}\times N$ and the fluctuation should vanish. The value of $\bar{n_\mathrm{r}}$ is slightly bigger than the values reported in Refs. \cite{Sun08,Olmos09}. As we will see in the next section this is due to the particular choice of the initial state.

The distribution $\rho_m^\mathrm{steady}(\alpha)$ contains the full statistics of the Rydberg atom number count. Strong interactions are known to have an effect on the counting statistics leading to a sub-Poissonian distribution \cite{CubelLiebisch05,Ates06} of the Rydberg number. A measure for this is given by the Mandel Q-parameter
\begin{eqnarray}
  Q=\frac{\bar{n^2_r}-\bar{n_r}^2}{\bar{n_r}}-1,
\end{eqnarray}
which is negative/positive for a sub-/super-Poissonian distribution of the Rydberg atom number count. In the steady state we find
\begin{eqnarray}
  Q_\mathrm{steady}=\frac{N\,\sigma^2_\alpha}{\alpha_\mathrm{max}}-1=\frac{\sqrt{5}-9}{10}\approx -0.676
\end{eqnarray}
which shows the expected sub-Poissonian behavior.

For the sake of completeness let us consider the case of non-interacting atoms. Here one obtains for the probability density
\begin{eqnarray}
  \rho_m(t)=\left(
              \begin{array}{c}
                N \\
                m \\
              \end{array}
            \right)\sin^{2N}t \cos^{2N-2m} t\label{eq:rho_non_interacting}
\end{eqnarray}
and hence the probability density in excitation number space performs an oscillatory motion at all times.

\section{Numerical results}\label{txt:numerics}
We are now going to compare the results that we obtained from the previous section to the actual data obtained from a numerical propagation of the Schr\"odinger equation. In order to make the numerical solution feasible we massively exploit the symmetry properties of the system. In all numerical calculations we refer to a set of basis states which are invariant under cyclic shifts and reversal of the lattice sites as has been used in Ref. \cite{Olmos09}. I.e. we operate only in a subspace of the space which is spanned by all states being compatible with the perfect blockade condition (\ref{eq:non_interacting_condition}).

\subsection{Evolution into the steady state}
\begin{figure}\center
\includegraphics[width=.9\columnwidth]{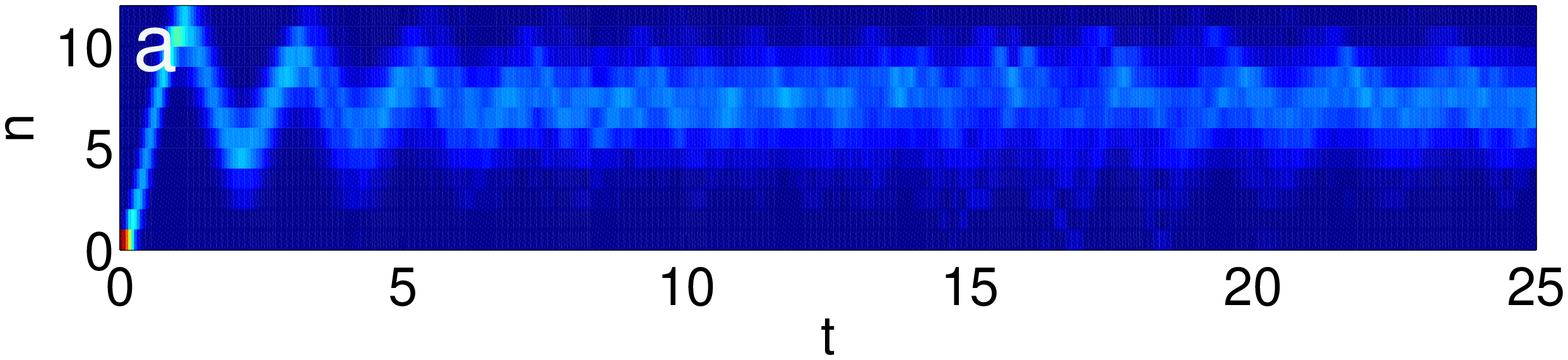}
\includegraphics[width=.9\columnwidth]{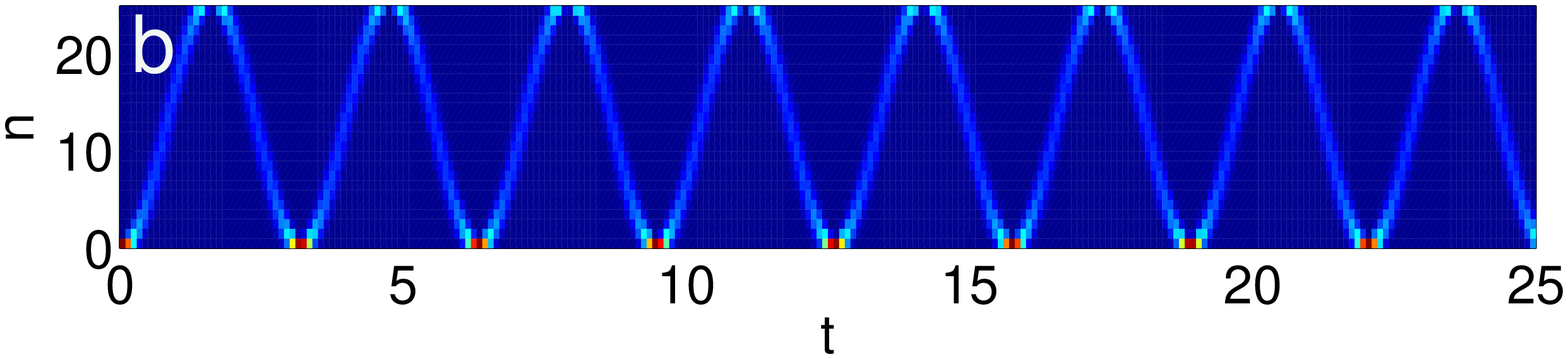}
\caption{Temporal evolution of the system consisting of 25 atoms in particle number space. Initially all atoms are in the ground state, i.e. $\rho_0=1$. \textbf{a}: In the interacting case (perfect blockade) the system reaches eventually a state in which the probability density localizes in excitation number space. \textbf{b}: This is not the case in the absence of interactions. Here, the wave packet performs coherent oscillations with maximal amplitude. Note the different scale of the $n$-axis.}\label{fig:evolution_of_rho}
\end{figure}
\begin{figure}\center
\includegraphics[width=0.5
\columnwidth]{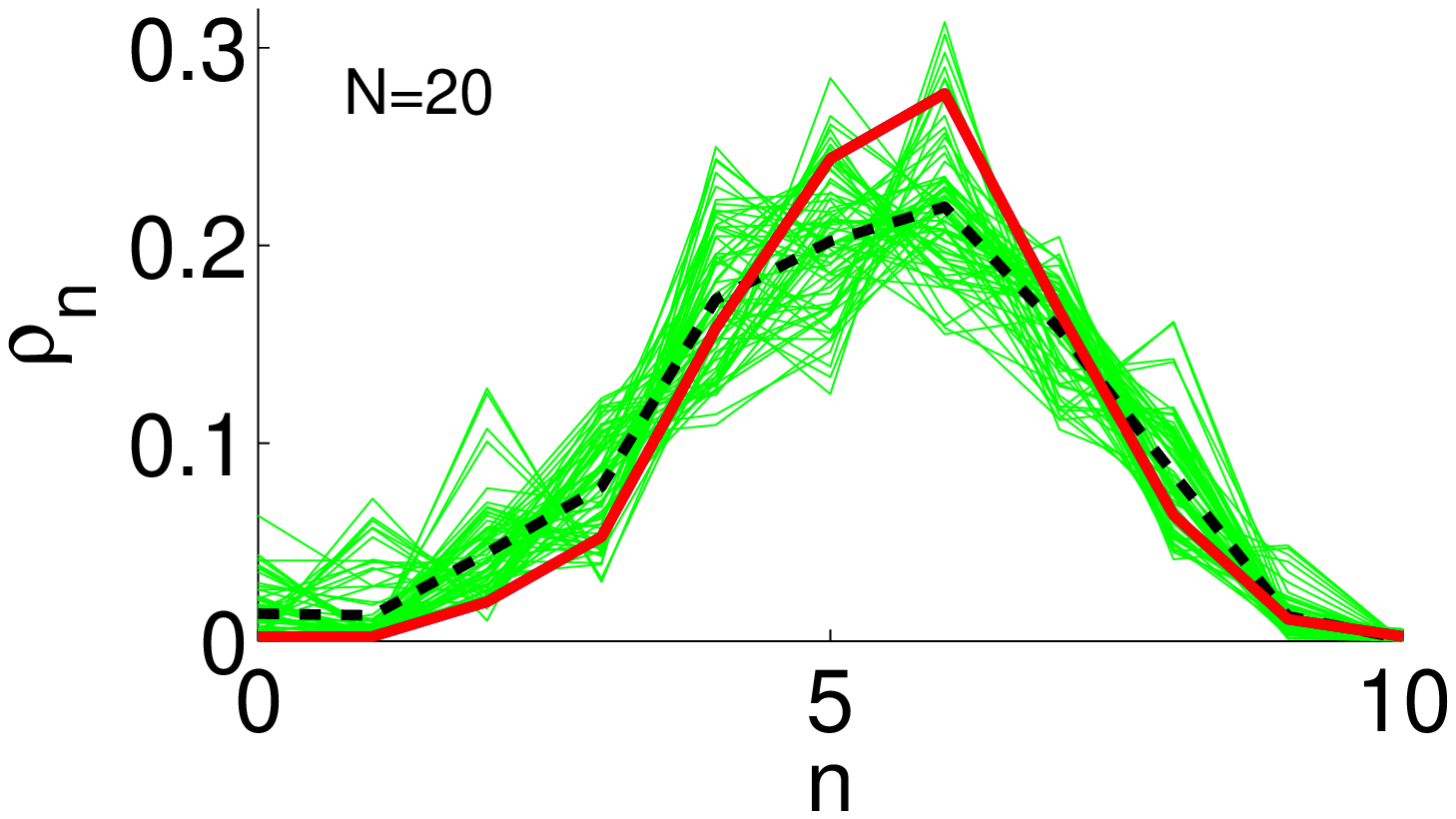}\includegraphics[width=0.5
\columnwidth]{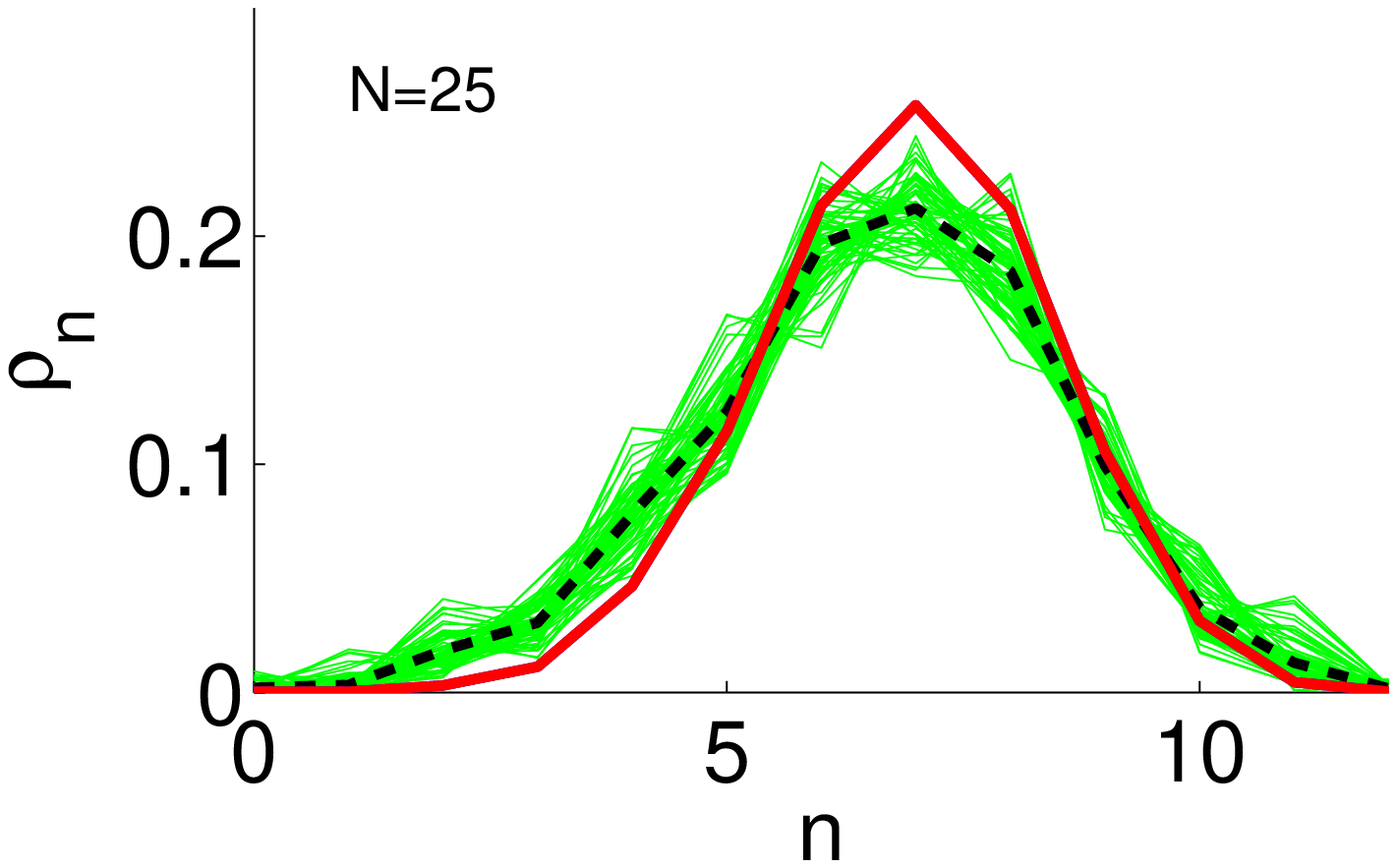}
\caption{Probability density $\rho_n$ in excitation number space for $N=20$ and $N=25$. The green (thin) curves are snapshots taken during the interval $100\leq t \leq 104$. For these times the calculated Rydberg number shows the steady state shown in fig. \ref{fig:generic_rydberg_number}. The dashed curve is obtained by taking the average over the set of snapshots. The red curve shows $\rho_n^\mathrm{steady}$ as given by eq. (\ref{eq:steady_state_condition}).}\label{fig:distribution_plots}
\end{figure}
Let us start by inspecting the evolution of $\rho_n$ when choosing the vacuum as initial state, i.e. $\rho_0=1$, and $N=25$. The result is shown in fig. \ref{fig:evolution_of_rho}a. For $t\leq 5$ we observe a well-defined wave packet which propagates through the excitation number space performing an oscillatory motion. For longer times the amplitude of the oscillations decreases, however, the wave packet remains localized. There are still significant fluctuations visible. In particular in the interval $15\leq t\leq 25$ remnants of the initial oscillations can be observed. In comparison to the non-interacting case which is shown in fig. \ref{fig:evolution_of_rho}b the localization in excitation number space is apparent. As expected from eq. (\ref{eq:rho_non_interacting}) here the wave packet exhibits coherent oscillations with large amplitude.

\subsection{The steady state and its dependence on the initial condition}
We now proceed by monitoring $\rho_n(t)$ over a time-interval in which the number of Rydberg atoms shows the steady state behavior -  here we choose $100\leq t \leq 104$. The result is depicted in fig. \ref{fig:distribution_plots} for two values of $N$, $20$ and $25$. The thin green curves show individual snapshots of $\rho_n(t)$ taken at different times. In addition we present also the average of $\rho_n(t)$ taken over the considered time-interval. The fluctuations around this average decrease significantly with increasing $N$. This is in accordance with the behavior of the Rydberg number $n_\mathrm{r}$ whose fluctuations around the mean value also diminish as $N$ increases (see Ref. \cite{Olmos09}). This supports the assumptions that in the limit of very large $N$ indeed a steady state with extremely little fluctuations is established. For both values of $N$ shown in fig. \ref{fig:distribution_plots} a comparison to the steady state result (\ref{eq:equilibrium_state}) (thick red curve) reveals a shift of the probability distribution to smaller $n$. These deviations appear to stem from the particular choice of the initial state: The state $\left|\Psi(0)\right>=\left|0\right>$ is localized at the leftmost vertex of the network. In this region of the graph the matrices $\mathcal{C}_{n,n+1}$ are, however, not actually random since the 'randomness' is caused by the interaction which has little or no effect when the number of Rydberg atoms is only very small, i.e. $n=0,1,2$.
\begin{figure}\center
\includegraphics[width=0.5
\columnwidth]{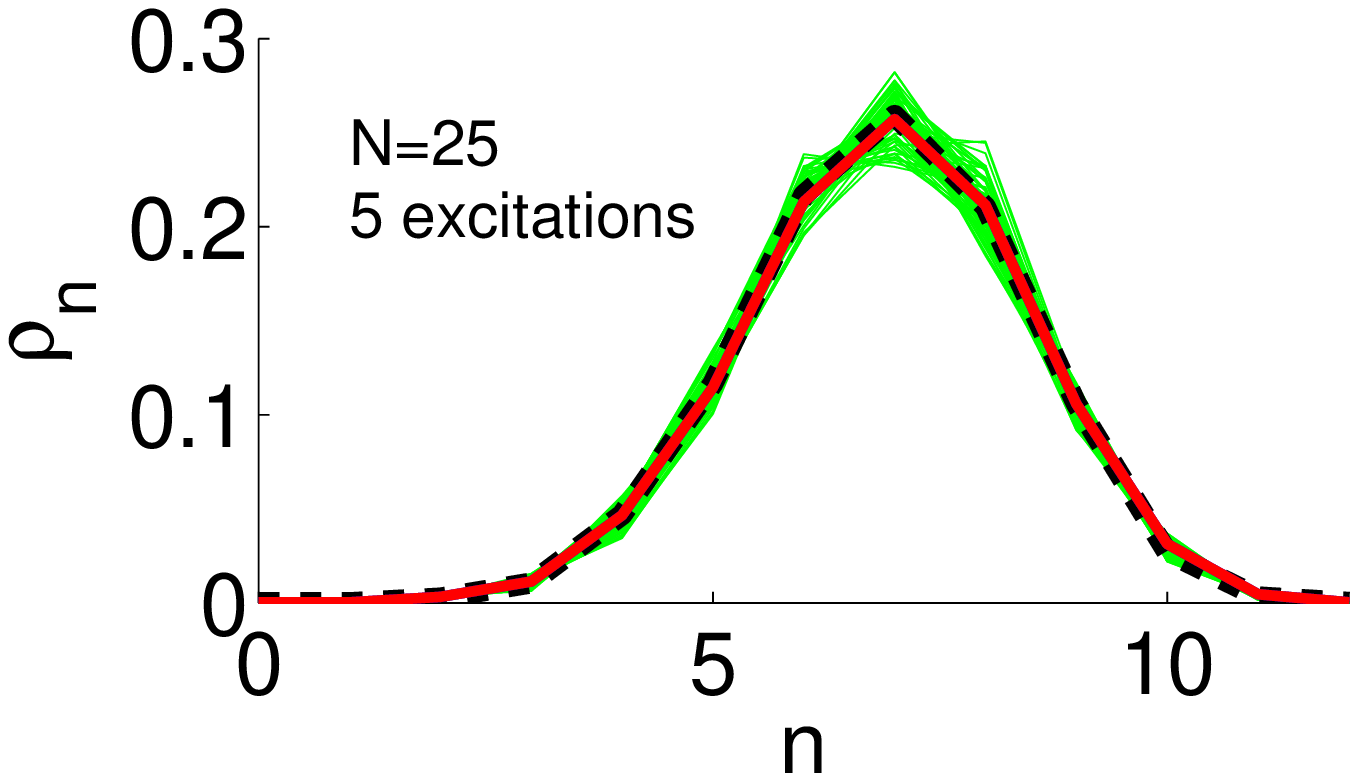}\includegraphics[width=0.5
\columnwidth]{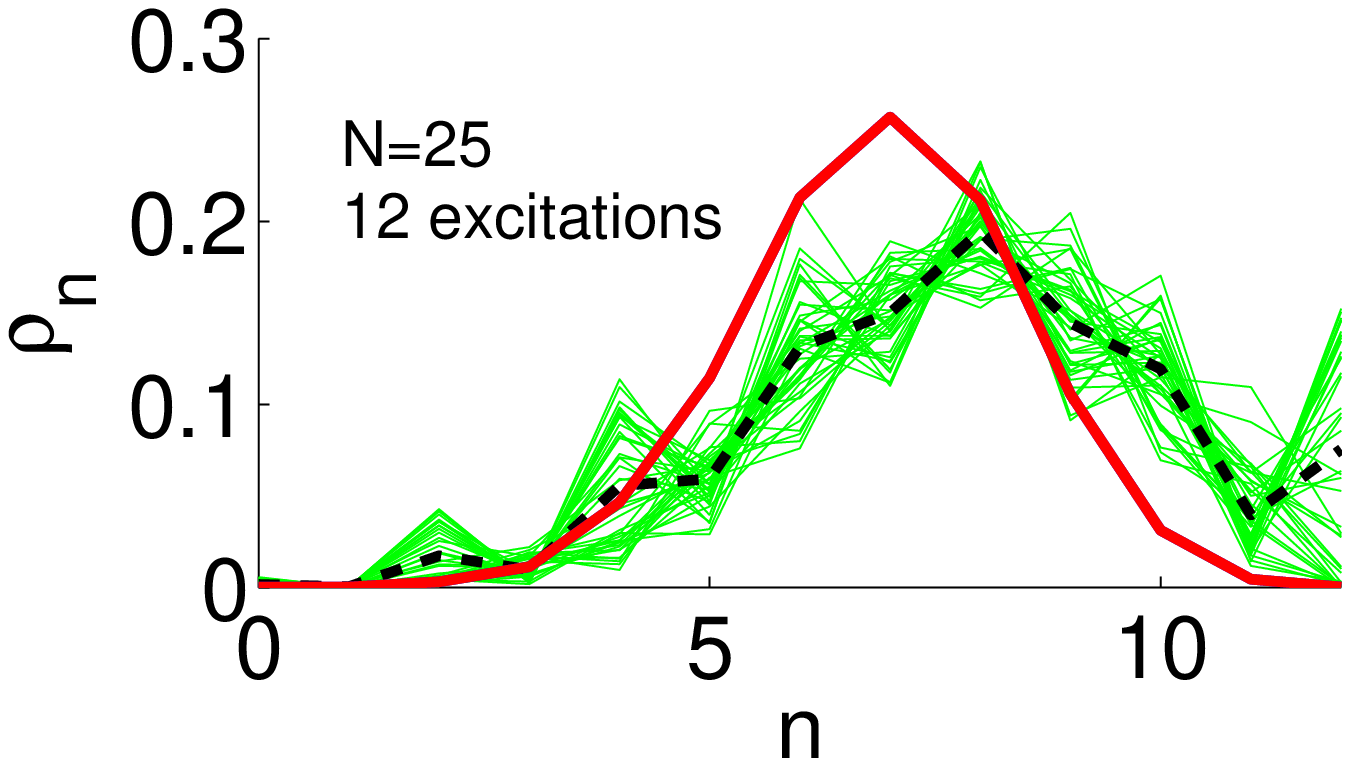}
\caption{The same plot as in fig. \ref{fig:distribution_plots} using a state containing $5$ and $12$ excitations as initial state. In the former case the fluctuations around the average value are small and the agreement with $\rho_n^\mathrm{steady}$ is remarkable. The latter shows substantial deviations from the steady state (\ref{eq:equilibrium_state}).}\label{fig:distribution_plots_not_vac}
\end{figure}
That this 'edge effect' appears to be indeed the cause of the deviation of the probability distribution from $\rho_n^\mathrm{steady}$ is corroborated by the data shown in fig. \ref{fig:distribution_plots_not_vac}. Here we present the same plot as in fig. \ref{fig:distribution_plots} but the initial state has to be chosen from the subspace containing 5 excitations, e.g. it is located in the central region of the graph. The effect is not only a much better agreement of the data with $\rho_n^\mathrm{steady}$ but also a significant decrease of the fluctuations about the average of $\rho_n(t)$. This behavior is generic for initial states chosen from excitation number subspaces with large dimensions.

Large deviations of the steady state from eq. (\ref{eq:equilibrium_state}) are again encountered when the initial state is located close to the right hand edge of the graph, i.e. when its number of Rydberg atoms is close to $n_\mathrm{max}$ (see fig. \ref{fig:distribution_plots_not_vac}). Hence, it becomes evident that eq. (\ref{eq:final_equation_for_rho}) is not unconditionally valid. It constitutes a reliable approximation only if the initial state belongs to a particle number subspace of sufficiently large dimension. That it also works well, when the vacuum is chosen as the initial state is not evident.

\subsection{Reduced density matrix in excitation number space}
\begin{figure}\center
\includegraphics[width=0.5
\columnwidth]{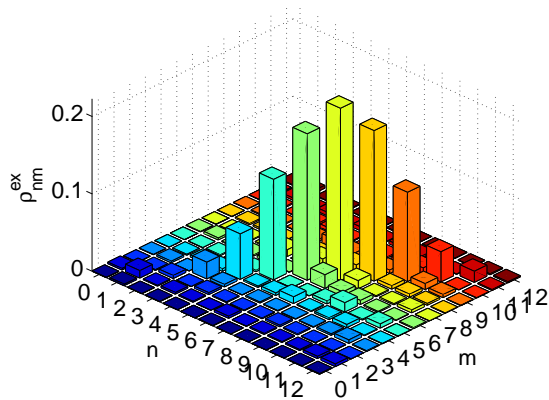}
\caption{Reduced density matrix in excitation number space $\rho^\mathrm{ex}$ at $t=140$. ($N=25$, initial state $\left|0\right>$). Shown is the absolute value of the entries. $\rho^\mathrm{ex}$ is well-approximated by a completely mixed state. }\label{fig:reduced_density_matrix}
\end{figure}
So far we have only studied the probability density distribution in excitation number space. Further insights can be gained by examining the reduced density matrix in excitation number space $\rho^{\mathrm{ex}}$ as this quantity eventually determines the outcome of a measurement of the number of Rydberg atoms. Its elements are defined by
\begin{eqnarray}
  \rho^\mathrm{ex}_{nm}=\sum_{\alpha=1}^{\mathrm{min}(\mathrm{dim}_m,\mathrm{dim}_n)} \left[\mathbf{\Psi}_n \otimes\mathbf{\Psi}^\dagger_m\right]_{\alpha\alpha}.
\end{eqnarray}
In fig. \ref{fig:reduced_density_matrix} we present a snapshot of $|\rho^{\mathrm{ex}}|$ for a system of $25$ sites at the time $t=140$. The initial state was $\left|0\right>$. We clearly observe that the entries of the main diagonal dominate the off-diagonal entries. Hence, there is negligible coherence between excitation number subspaces which have a large dimension. Consequently, the reduced density matrix is well approximated by a classical mixture $\rho^\mathrm{ex}\approx \sum_{n=0}^{n_\mathrm{max}}\rho_n \left|n\right>\left<n\right|$. The strong interaction between the atoms in conjunction with the laser driving erases the phase relation between excitation number subspaces. So tracing out all degrees of freedom but those being relevant for the measurement of the Rydberg number, leaves us with a density matrix of a completely mixed state. This gives actually the impression that the steady state we observe is a state with maximal entropy, since only the dimension of the excitation number subspaces determines the outcome of the measurement (see eq. (\ref{eq:equilibrium_state})). In fact we are dealing with a pure state at all times and only the particular measurement we are performing gives us the impression of observing a completely mixed state.

\subsection{Connection with the microcanonical ensemble}
Note that all the information about the steady state could as well have been obtained by considering a microcanonical ensemble. Here, the microstates are just given by the zero energy eigenstates $\{\phi\}$ of the interaction Hamiltonian $H_\mathrm{int}$ defined through condition (\ref{eq:non_interacting_condition})) and the steady state given in eq. (\ref{eq:rho_analytical}) can be obtained directly by counting the number of these microstates.
The fundamental assumption underlying the microcanonical ensemble is that each of the microstates has equal weight. This assumption can clearly not be justified in the absence of the laser, even though all states have the same energy. Only when the laser is present, the microstates $\{\phi\}$ defined by (\ref{eq:non_interacting_condition})) become strongly mixed and are thus no longer eigenstates of the system. However, they no longer possess strictly zero energy but are rather distributed over an energy interval which is centered at zero and whose width is proportional to $N\Omega$. In other words, although the laser produces a widening of the energy window occupied by the states, it also provides the necessary ingredient that eventually allows the system to thermalize, the equiprobability of the microstates.

At this point we want to remark that the microcanonical prediction to the steady state (\ref{eq:rho_analytical}) involves all zero energy eigenstates of $H_\mathrm{int}$. On the other hand, the numerical results of section \ref{txt:numerics} involve only a subset of all accessible states, i.e. the fully symmetric set of eigenstates. The observed agreement between the two approaches suggests that the number of fully symmetric eigenstates with a given excitation number $m$ is proportional to $\mathrm{dim}_m$.

\section{Summary and conclusions}
We have investigated the origin of the steady state value of the Rydberg number which is exhibited in a laser driven Rydberg gas after an initial transient period. Starting from Heisenberg's equation we have derived an effective equation of motion for the probability density in excitation number space. This effective equation of motion which is coarse-grained in time exhibits a steady state. When comparing this steady state to actual numerical simulations excellent agreement is found provided that the initial state was chosen from an excitation subspace with sufficiently large dimension. In case of an initial state containing a very small/large number of excitations still a steady state is established, however, deviations from the analytical result are obtained.

We have visualized the system by a graph whose vertices are represented by eigenstates of the interatomic interaction. Coupling between the vertices is established by the laser-atom interaction. A similar mapping was applied in Refs. \cite{Altshuler97,Berkovits98} where interacting fermions were studied by means of a graph. Here, a transition between localized and delocalized eigenstates has been found to take place as a function of the interaction strength. In our system we are in the regime of strong interaction and the eigenstates are delocalized throughout the entire graph. The observed localization in excitation number space and hence also the observation of a steady state value of $n_\mathrm{r}$ is a purely statistical effect, owed to the strongly peaked function $\mathrm{dim}_m$.

It would be interesting to see whether a distribution of the Rydberg number count similar to eq. (\ref{eq:equilibrium_state}) and - more specifically - the calculated values for the mean Rydberg number and the Mandel Q-parameter can be observed in actual experiments. Studying the shape and the temporal evolution of the distribution should yield insights into how this steady state is established as the interaction strength increases. Since our simple model is not expected to be valid in higher dimension experiments with Rydberg atoms in lattices could also help here to clarify whether and, if so, how a steady state is established.

\ack
Discussions with R. Gonz\'{a}lez-F\'{e}rez are gratefully acknowledged. B.O. acknowledges the support of Ministerio de Educaci\'{o}n y Ciencia under the program FPU. B.O. also gratefully acknowledges the MICINN grant FIS2008-02380
and the JA grant 2445.
\\
\bibliographystyle{iopart-num}

\end{document}